This is a preprint of the following chapter: Riordan, N., Machoň, M., Csajková, L.K., Space Diplomacy Toward a Legal Regime for Space Resource Activities on the Moon, published in Legal Responses to Space Resource Utilisation. Space Law and Policy, edited by Bhat B, S., Variath, A., 2026, Springer, Singapore. It is the version of the author's manuscript prior to acceptance for publication and has not undergone editorial and/or peer review on behalf of the Publisher (where applicable). The final authenticated version is available online at: https://doi.org/10.1007/978-981-95-8864-0_7

# SPACE DIPLOMACY TOWARD A LEGAL REGIME FOR SPACE RESOURCE ACTIVITIES ON THE MOON

Nancy Riordan[1]

Miloslav Machoň[2]

Lucia Kozák Csajková[3]

**Abstract:**

The development of national policy and legal outputs toward space resource utilization across several countries accelerated after the introduction of the Artemis Accords in 2020. The announcement also further intensified the debate about the necessity of specifying a regime for using space resources on the international level. In 2021 the Legal Subcommittee of the UN Committee on the Peaceful Uses of Outer Space finally approved the establishment of its Working Group on Legal Aspects of Space Resource Activities. Therefore, this chapter builds on diplomacy under conditions of polycentric governance and aims to map the diversity of actors' ideas and interests about governance for space resource utilization on the level of the Working Group. The research process follows qualitative analysis methodology and assigns the results into the matrix of prospective alternatives for governing space resources. The available materials of the Working Group including reports and recordings, constitute the core empirical data used in the analysis.

[1] University of Massachusetts, Boston

[2] University College Prague

[3] Prague University of Economics and Business

## Introduction

The Working Group on the Legal Aspects of Space Resource Activities (hereafter as the Working Group) was established under the UN Committee on the Peaceful Uses of Outer Space (UNCOPUOS) Legal Subcommittee agenda item "General exchange of views on potential legal models for activities in the exploration, exploitation and utilization of space resources" at the sixtieth session of the Legal Subcommittee in 2021, and later renamed in 2022.[4] The establishment of the formal working group was preceded by years of informal international discussions and permissive regulatory measures taken at the national level by a few states, beginning with authorizing commercial activities in U.S. legislation.

While commercial efforts in space had existed, they were significantly supported with the 2010 NASA Authorization Act. It called for an extension of space exploration for which space resource utilization would be needed.[5] Also, it anticipated active U.S. involvement and leadership in commercial space activities presuming to support commercial endeavors. Within this Act, the U.S. strived for research, development, space exploration and international cooperation in the field of space exploration,[6] as well as maintaining safe and affordable space exploration capabilities throughout the next century.[7] These assumptions for space resource utilization were formalized in the U.S. Commercial Space Launch Competitiveness Act of 2015. It elevated commercial activities in space to "promote the right of United States citizens to engage in commercial exploration for and commercial recovery of space resources free from harmful interference, in accordance with the international obligations of the United States and subject to authorization and continuing supervision by the Federal Government".[8] However, the U.S. Act ruled out asserting sovereignty, ownership or other exclusive rights over any celestial body.[9]

---

4 UNOOSA, "Working Group on Space Resources."

5 Space Review, "The Space Review: Putting the White House Executive Order on Space Resources in an International Context."

6 Robinson, "Statement of The Honorable Elizabeth M. Robinson Chief Financial Officer National Aeronautics and Space Administration before the Committee on Commerce, Science and Transportation United States Senate."

7 Robinson.

8 Sen. Rockefeller, "S.3729 - 111th Congress (2009-2010)."

9 Sen. Rockefeller.

In addition to the U.S., other countries expressed a strong interest in the commercial exploitation of space resources. The Luxembourg Space Resources Initiative was formed and later supported by the Law of 20 July 2017 on the exploration and use of space resources even before Luxembourg established its own Space Agency in 2018.[10] The Luxembourg initiative advocates that space resources should be used for peaceful purposes for the common benefit of humankind, and are gathered sustainability and in compliance with international law, such as the Outer Space Treaty.[11] According to this initiative, the resources from Near Earth Objects should be explored and utilized commercially and indicates that an appropriate legal and regulatory framework for space resource utilization activities should be established.[12] Moreover, the program supported the relevance of polycentric governance[13] and to connect diverse public and private organizations to aid in space exploration - including Luxembourg Government and Ministry of economy, as well as U.S. and Japanese private companies.

In response to these developments and the intensifying debate on the commercial use of space resources, in 2017, upon a proposal by the Belgian delegation, the topic of space resources first appeared as a single-year agenda item of the UNCOPUOS Legal Subcommittee. This became agenda item 14, General exchange of views on potential legal models for activities in exploration, exploitation and utilization of space resources, and received ever-increasing attention by the legal subcommittee.[14]

Furthermore, in 2016 the Hague International Space Resources Governance Working Group was formed, following ongoing informal discussions and a Roundtable on the Governance of Space Resources, convened by The Hague Institute for Global Justice on 1 December 2014. The group's responsibility was to ensure that the discussion about space resource activities were meeting the existing treaty obligations, regarding on-orbit operations and space resource rights. In November of 2019 the Hague Working Group adopted the Building Blocks for the Development of an International Framework on Space Resource Activities (hereafter as the Building Blocks). The Building Blocks have since served as a key reference for negotiations in the Working Group.[15]

---

[10] Journal officiel, "Journal Officiel Du Grand-Duché de Luxembourg: Loi Du 20 Juillet 2017 Sur l'exploration et l'u... - Legilux."

[11] SpaceResources.lu, "The SpaceResources.Lu Initiative Positions Luxembourg as a Pioneer in the Exploration and Utilisation of Space Resources."

[12] SpaceResources.lu, "SpaceResources.lu."

[13] Kim, "Is Global Governance Fragmented, Polycentric, or Complex?," 905–6.

[14] SpaceResources.lu, "The SpaceResources.Lu Initiative Positions Luxembourg as a Pioneer in the Exploration and Utilisation of Space Resources."; ISPACE.INC, "Lunar Exploration Company Ispace and Luxembourg Government Sign MoU to Co-Operate Within the SpaceResources.Lu Initiative"; gouvernement.lu, "SpaceResources.Lu: The Luxembourg Government Becomes a Key Shareholder of Planetary Resources, Inc., the U.S.-Based Asteroid Mining Company."

The commercial use of space resources became a more significant topic, and could be observed during the ongoing sessions of the UNCOPUOS Legal Subcommittee. By the fifty-seventh session of the Legal Subcommittee held in Vienna from 9 to 20 April 2018 the need to ensure sustainability of space activities was stressed by several delegations and that the LSC would require a new impulse to keep up with current trends of commercialization and privatization of outer space.[16]

Linking to the previous discussions, in the fifty-eighth session, Belgium and Greece introduced a working paper containing a proposal for working methods and a workplan for a new working group on legal aspects of the exploration, utilization and exploitation of space resources.[17] The initiative of Belgium and Greece and the following discussion on this topic was supported by various delegations. However, not all delegations supported the formation of such a group. According to the opposing views, the legal framework was sufficient in regard to space activities and the exploration, exploitation and utilization of space objects were still being innovated and forming a working group on legal aspects could hinder these innovations. If the working group were to be established, it should encourage public and private investment in this area, as well as respect individual rights and collective interest in the commercial exploitation of space objects. According to the members, it was essential that the discussion, in the potential working group, should also include scientific, economic, technical and legal experts.[18]

Reflecting on these developments in the international arena, by April 2020 the U.S. issued Executive Order 13914 "Encouraging International Support for the Recovery and Use of Space

---

[15] Universiteit Leiden, "The Hague International Space Resources Governance Working Group."

[16] UNOOSA, "A/AC.105/1177 - Report of the Legal Subcommittee on Its Fifty-Seventh Session, Held in Vienna from 9 to 20 April 2018."

[17] UNOOSA, "A/AC.105/C.2/2019/CRP.26 - Proposal for Working Methods and Work Plan of the Working Group on Legal Aspects of the Exploration, the Utilization and the Exploitation of Space Resources, (with Reference to Document A/AC.105/C.2/L.311): Working Paper by Belgium and Greece," 1–2.

[18] UNOOSA, "A/AC.105/1203 - Report of the Legal Subcommittee on Its Fifty-Eighth Session, Held in Vienna from 1 to 12 April 2019."

Resources." The commercial activities and utilization of lunar resources were backed by this new executive order that encouraged commercial partners to participate in the U.S. "innovative and sustainable program",[19] in order to return humans to the Moon and launch a mission to Mars. Although, according to this executive order, the U.S. would not view outer space as a global common, the Secretary of State should negotiate and arrange bilateral and multilateral agreements with foreign states.[20] This executive order would constitute a crucial part in achieving the objectives of the Artemis program and is one of the foundations for the future of a U.S. commercial space economy.[21] Later that year, in October 2020, the U.S. led Artemis Accords were announced.[22] The Artemis Accords reinforce that space resource extraction and utilization can and will be conducted under the auspices of the Outer Space Treaty, with specific emphasis on Articles II, VI, and XI.

The variety of national and international debates and initiatives accelerated and deepened after 2020 when U.S. representatives, along with partner countries, introduced the Artemis Accords.[23] Therefore, the aim of our research is to understand a diversity of actors' interests and emerging ideas about governing space resource utilization with an emphasis on the Moon in the Working Group. Understanding diplomatic negotiations under conditions of polycentrism[24] has the potential to deepen knowledge about the requirements for achieving a compromise between diverse actors with unique political demands, socio-economic social bases and cultures. It opens avenues for future theorizing about whose ideas about governance goals under schema for common pool resources that are as yet not in use as such (future) but are evolving and how.

The research methodology had three steps. First, we began with a narrative literature review approach intended for topics that have been conceptualized differently and studied by various groups of researchers within diverse disciplines[25] in order to detect relevant emerging themes, theoretical perspectives, and common issues.[26] A Web of Science advanced bibliometric search and analysis for 1984-2024 yielded 1,578 documents published in the Web of Science Core Collection using topic search terms "lunar resources." Of these, 998 sources were published between 2015-2024 with the most articles, book chapters, and other sources published in 2023. Further refining the process we also used the HeinOnline database (a collection of legal and other journals) using search terms "Lunar resource governance," yielding 448 results with the most research published on Lunar resource governance between 2020 and 2024.

---

[19] Federal Register, "Executive Order 13914: Encouraging International Support for the Recovery and Use of Space Resources."

[20] Federal Register.

[21] Space Review, "The Space Review: Putting the White House Executive Order on Space Resources in an International Context."

[22] Riordan, Machoň, and Csajková, "Space Diplomacy and the Artemis Accords," 19.

[23] Riordan, Machoň, and Csajková, 19.

[24] Dorsch and Flachsland, "A Polycentric Approach to Global Climate Governance," 53; Scholte, "From Government to Governance," 56.

Second, the results of literature review highlighted the space diplomacy mechanisms conceptualized by Cross and Pekkanen[27] and model for governing space resources introduced by Yap et al.[28] to systematize trends in starting diplomatic negotiations at the Working Group. The space diplomacy conceptualization outlined communication, persuasion, and bargaining as three common mechanisms for diplomatic activities of governmental and non-governmental actors. The matrix model for governing space commons by Yap et al. recognized two dimensions. The first dimension covered the distribution and resource access; the second one dealt with the Earth-bound sustainability. It stemmed from global governance studies elaborated aspects of non-rivalry and non-excludability in IR[29], future studies conceptualized and promoted scenario models for predictions[30], and science and technology studies connected the scenario analysis with a socio-technological nexus.[31]

Third, a qualitative analysis strategy constituted the core of used methodologies. Applying the rules of systematic sampling,[32] we compiled submitted materials from countries and observer organizations 2022-2023 publicly available from the Working Group, including reports and recordings. In the first phase, qualitative content analysis followed the rules of structured coding to sort the collected data into two categories: government inputs and observer (non-government) inputs to the Working Group.[33] The pattern coding, which is the analogy for cluster analysis used in quantitative fields, constituted the second cycle of the qualitative analysis. We then combined the summaries from the previous cycle into the four categories developed by Yap et al.[34] This process revealed the meaning within the data and integrated this into higher-order key themes to explain initial codes in a more general and thematic way.[35]

---

[25] Wong et al., "RAMESES Publication Standards," 2.
[26] Snyder, "Literature Review as a Research Methodology," 335.
[27] Cross and Pekkanen, "Introduction. Space Diplomacy," 202–6.
[28] Yap et al., "Four Alternative Scenarios of Commons in Space," 397–402.
[29] Kaul and Mendoza, "Advancing the Concept of Public Goods," 82–83.
[30] Oliver Schwarz, "Assessing the Future of Futures Studies in Management," 238–39.
[31] Jasanoff, "One. Future Imperfect," 7–9.
[32] Krippendorff, *Content Analysis*, 118–19.

The chapter has six sections. It begins with a theoretical discussion about polycentricity followed by its manifestations in space diplomacy. The second section reviews a selection of existing literature on space resource utilization and governance. The third section provides operationalization of the Yap et al., model including four scenarios for global common governance in the context of space diplomacy strategies. The fourth section provides results of the qualitative analysis. Fifth, the discussion confronts the results with the latest findings on space resource utilization identified in the literature review. The conclusion offers suggestions for future research.

## Toward Polycentric Governance

In the 1950s, Michael Polanyi dealt with managing polycentric tasks for which he emphasized an element of mutual harmonization of individual activities.[36] Following his works, Ostrom elaborated his concepts and labeled polycentricity a useful analytical approach for understanding climate change governance, because it can capture diverse actors' actions participating in natural resource, environmental, and climate governance at different scales.[37] In the 1990s, polycentricity was adopted in International Relations (IR) for studying the sustainability of common-pool resources under conditions of global interdependence[38], because the globalization processes significantly blurred the boundaries between the foreign actions of governmental and external activities of non-governmental actors and deepened interdependence between them.[39]

[33] Saldaña, *The Coding Manual for Qualitative Researchers*, 10; Gibbs, *Analyzing Qualitative Data*, 55–57.
[34] Saldaña, *The Coding Manual for Qualitative Researchers*, 238.
[35] Ritchie et al., *Qualitative Research Practice*, 271; Krippendorff, *Content Analysis*, 213.
[36] Polanyi, *The Logic of Liberty*, 184.
[37] Dorsch and Flachsland, "A Polycentric Approach to Global Climate Governance," 47.
[38] Keohane and Ostrom, *Local Commons and Global Interdependence*, 3–5.
[39] Bolewski, "Compass for Public/Private Management in Turbulent Times," 8.

Polycentricity admits adjusting and aligning toward a collective goal in conditions of decentralism.[40] Therefore, the polycentric approach presumes the fragmentation of state and competitive growth of public-private modes of governance with larger and smaller jurisdictions than a country domain.[41]

In 2012, Aligica and Tarko conceptualized monocentrism as the opposite of the polycentric mode of governance.[42] These modes of governance differ in two fundamental aspects.[43] First, centralization is a prerequisite of monocentric policy processes, whereas polycentrism allows a far more dispersed formulation, implementation, and enforcement of social rules to reduce an individual failure.[44] Second, the monocentric mode of governance includes only government as the relevant actor, while polycentric governance invites the empirical diversity of global governance. It broadens the analytical scope to capture various kinds of actors and their diverse interactions. Theorizing about the diversity of interactions distinguishes local governments, national governments, regional institutions, transnational companies, and distinguished individuals.[45] These diverse actors with local, national, regional, and global jurisdictions carry out societal regulation across public and private sectors, including their hybrid public-private forms.[46]

From an analytical perspective, monocentric governance primarily focuses on a country level, whereas polycentric governance recognizes several spatial scales.[47] While monocentrism emphasizes neatly separated territorial jurisdictions, polycentrism often involves overlapping jurisdictions. Following empirical results, considering the variety of spatial configurations allows tackling diverse problems on spatial scales effectively, supporting the legitimacy of decision-making processes and their outputs.[48] A distinction also lies in allowable modes of regulatory control. While supreme, comprehensive, and exclusive regulatory control is acceptable for monocentrism, polycentrism labels this form of control as undesirable and prioritizes conditions in which no actor of governance can aspire for a predominant position. Research on polycentricity focuses on autonomous external policies and activities of sub governmental actors to highlight their foreign policy-making competencies.[49]

---

40 Kim, "Is Global Governance Fragmented, Polycentric, or Complex?," 905–6.
41 Scholte, "From Government to Governance," 67.
42 Aligica and Tarko, "Polycentricity," 244.
43 Scholte, "From Government to Governance," 49.
44 Alshamy et al., "Polycentric Defense, Ukraine Style," 40.
45 Baldwin et al., "Empirical Research on Polycentric Governance," 322–23.
46 Dorsch and Flachsland, "A Polycentric Approach to Global Climate Governance," 50.
47 Scholte, "From Government to Governance," 49.
48 Newig, Schulz, and Jager, "Disentangling Puzzles of Spatial Scales and Participation in Environmental Governance—The Case of Governance Re-Scaling Through the European Water Framework Directive," 1000–1001.

In the context of IR, polycentrism opens avenues for studying the self-organization and coordinated networking of these diverse actors in policy arenas in specific contexts.[50] The self-organization principle claims the nondominant position of an individual or a group of actors and secures the freedom to set up their own rules. The site-specific context - i.e. a multilateral institution such as the UN Committee on the Peaceful Uses of Outer Space (COPUOS) and resources in ABNJ (areas beyond national jurisdiction) implies a polycentric task for an individual or a group of actors. These tendencies highlight the relevance of global governance networks in which institutional ties and collaboration accelerate negotiations among diverse actors in international policy arenas.[51] Understanding diplomatic negotiations under polycentrism has the potential to deepen knowledge about the requirements for achieving a compromise between diverse actors with unique socio-economic social bases, political interests and cultures.[52] It may also clarify conditions for achieving governance goals under schema for common pool resources - that are as yet not in use as such (future).[53]

*Specifics Of Diplomacy Under Conditions of Polycentrism*

Whereas, under conditions of monocentrism, diplomacy has always involved a diplomat who exclusively represented one nation-state in the territorial jurisdictions of another state, in the case of public diplomacy, polycentrism recognizes separate governance agencies with diversity in their missions and the responsibility to represent a government in foreign territories.[54] The prerequisite for a diversity of actors under conditions of polycentrism stems from democratization trends.[55] In democratic regimes, national governments usually support a variety of sub governmental and transgovernmental actors for greater political autonomy to supplement governmental roles, covering para-diplomatic and external relations agendas.[56]

[49] Reinsberg and Dellepiane, "The Domestic Sources of Sub-State Foreign Policymaking," 927–28.
[50] Morrison et al., "Building Blocks of Polycentric Governance," 486–89.
[51] Kim, "Is Global Governance Fragmented, Polycentric, or Complex?," 909.
[52] Smith, "Innovating for the Global Commons," 56.
[53] Dorsch and Flachsland, "A Polycentric Approach to Global Climate Governance," 53.
[54] Scholte, "From Government to Governance," 56.

These diverse actors with ambitions to participate in space resource utilization have acquired access to evolving multi stakeholder diplomatic arenas to provide public goods and generate soft power. Therefore, they became diplomatic co-actors under conditions of polycentrism.[57] In doing so, they have developed specialized departments in their organizational structures to deal with external relations, which are equivalent to governmental foreign affairs ministries. The newly created departments acquired internal responsibility for negotiations with external actors to obtain access to policy arenas and markets and intensify relations with governments and other actors in global civic society through partnerships and strategic coordination or acting as independent actors.[58]

The integration of diplomatic practices into the portfolio of their strategic planning activities helped to increase the predictability of their operations and interact with external actors in the complex and volatile global environment.[59] These new actors in diplomatic negotiations gained an opportunity to shape human rights, improve global economic stability, promote global peace, and resolve local social conflicts.[60] They represent diverse overlapping and interweaving collective identities, including substate authorities, communities, and distinguished individuals.[61] Accompanying governmental officials at their diplomatic visits enables forming partnerships and coordinating activities with the officials.[62] These trends spill over into communication styles. Under polycentric governance conditions, a more diverse group of actors has the ability to communicate messages externally.[63] Therefore, diplomatic negotiations became more complex due to the larger number of diverse actors involved in diplomacy, which also increases aspects of cultural diversity.

[55] Bloodgood, "New Multilateralism: The United Nations and Governance in the Era of Nonstate Actors," 59, 61.
[56] Leffel, "Principles of Modern City Diplomacy and the Expanding Role of Cities in Foreign Policy," 183–84.
[57] Bolewski, "Compass for Public/Private Management in Turbulent Times," 8.
[58] Guilbaud, "Diplomacy by Non-State Actors," 189; White and Bolewski, "Corporate Diplomacy," 62–63.
[59] Ruël, "Multinational Corporations as Diplomatic Actors," 2.
[60] Ruël, 2.
[61] Jackson, "Paradiplomacy and Political Geography," 5.
[62] White and Bolewski, "Corporate Diplomacy," 64–67.

Regarding the Working Group participants in particular, these new actors can provide consultations in preparing multilateral negotiations or specialized conferences to articulate their interests and protect the existing legal environment. Moreover, if their activities in the global environment aim to provide economically and socially sustainable solutions, they have the potential to increase the legitimacy and trust of their activities for global society and enhance the image of the country of their origin.[64] Ensuring social and economic responsibility became an internally organizational form of governance for addressing sustainability issues. These new actors developed internal mechanisms for monitoring compliance with sustainability principles themselves, like codes of conduct.[65]

According to Cashore et al., interactions between traditional diplomats and new actors can take three forms: complementarity, competition, and coexistence. They act complementarily when they combine resources and talents to minimize transaction costs and mutually reinforce the same demand in negotiations[66]. The situation could be classified as competition when an antagonism for limited resources exists between traditional and new diplomatic actors aiming to monopolize power over these resources. Coexistence exists when traditional and new diplomatic actors cohabitate in an issue area, but traditional diplomats apply divergent goals and modes of governance.[67] Even if the actors in co-existential settings can act simultaneously, the coexistence may not be classified as harmonious.[68] In cases of the unpopularity of governments in foreign territories, these new actors have the potential to serve as balancers in public diplomacy.[69]

---

[63] Scholte, "From Government to Governance," 56.
[64] Bolewski, "Compass for Public/Private Management in Turbulent Times," 144.
[65] Renckens, *Private Governance and Public Authority*, 1–3.
[66] Milstein, "Capitalism and Democracy," 9.
[67] Cashore et al., "Private Authority and Public Policy Interactions in Global Context," 8–10.
[68] Milstein, "Capitalism and Democracy," 19.
[69] White and Bolewski, "Corporate Diplomacy," 63–67.

## Literature Review

One recurring theme across the last two decades of research on lunar governance involves complementary and contrasting responsibilities of governments and private actors in using space resources, usually resulting in policy recommendations for adaptive governance.[70] A previous study introduced structuring data on public-private interactions through scenarios to effectively display the advantages and disadvantages of each form of space governance effectively.[71] These authors also raised a dilemma between naturalist and positivist legal perspectives on whether existing international legal principles for property rights could be applied to space resources or not. Theorizing on historical experiences with human settlements of remote regions to build alternative predictions for lunar governance has gained popularity in scholarly journals across social science disciplines.[72]

Also, common among the literature is a growing recognition of the need for international cooperation to manage space resources effectively and sustainably.[73] Some authors provided comprehensive interpretations of the common heritage principle,[74] reflecting the recent needs of actors interested in lunar activities. The theorizing included correlation of the non-appropriation and common heritage principles used in the Outer Space Treaty and the Moon Agreement as legal concepts for setting relationships between spacefaring and non-space faring countries.[75] Studies also used the similarities and differences in international legal regimes for evaluation and comparison of domestic legislation regulating the use of outer space natural resources.[76] Moreover, some authors used the current state of the international legal regime for proposing prospective legal innovations for lunar resource utilization mining like environmental impact assessments[77] and priority rights for natural resources in space.[78]

---

70 Viikari, "Natural Resources of the Moon and Legal Regulation"; Masson-Zwaan and Sundahl, "National and International Norms towards the Governance of Commercial Space Resource Activity."

71 Jakhu and Buzdugan, "Development of the Natural Resources of the Moon and Other Celestial Bodies," 206, 228.

72 Moltz, "Toward Cooperation or Conflict on the Moon?," 89–90; Mouat et al., "(Un)Earthly Governance," 152–53.

73 Moltz, "Toward Cooperation or Conflict on the Moon?"; Hasin, "From 'Space Law' to 'Space Governance'"; Masson-Zwaan and Sundahl, "National and International Norms towards the Governance of Commercial Space Resource Activity"; de Zwart, Henderson, and Neumann, "Space Resource Activities and the Evolution of International Space Law."

74 Following recent discussion on preference of using gender-neutral terms, this chapter uses "common heritage of humankind" term to replace a well-established, but gender-biased terms "common heritage of humankind", "province of all mankind"Steer, "'The Province of All Humankind' – A Feminist Analysis of Space Law," 182–83..

75 Wang and Huang, "On the Common Heritage of Mankind Principle in Space," 230–32.

76 de Zwart, Henderson, and Neumann, "Space Resource Activities and the Evolution of International Space Law," 157–59; Hofmann, "Luxembourg Completing Its Space Legislation," 524–26.

Stakeholders and experts are increasingly advocating space governance approaches that provide flexible, policy-oriented frameworks to adapt to evolving technological capabilities.[79] They especially underline many unresolved issues in commercial lunar mining activities, as well as in the context of potential environmental degradation of these resources, for which the flexibility of governance mechanisms would be desirable. Besides the aspect of uncertainty of uses of natural space resources, Hasin[80] considered a space governance model, applicable also for lunar activities, based on the needs and desired outcomes of diverse actors as the most promising solution for achieving equal access to outer space and benefits from resource extraction.[81] Following the practical experiences in previous international policy and legal negotiations, some authors admitted that creating a multilateral regime for future space resource activities could be "a complex and lengthy process"[82] that will need considerable activity of non-governmental organizations to prepare an environment for reaching a consensus among countries. This process needs to include the diverse interests of different stakeholders, commercial actors in particular.[83] Experts close to the negotiations also found that "the future lunar legal landscape may well comprise international and national law evolving in parallel, at least for the near future".[84]

## Governing Space Resources

---

[77] Martin and Freeland, "Back to the Moon and Beyond," 428–29.

[78] Wang and Huang, "The Establishment of the Priority Right in Space Resources Activities," 920–23.

[79] Viikari, "Natural Resources of the Moon and Legal Regulation," 548–50; Neal et al., "The Moon Needs an International Lunar Resource Prospecting Campaign," 743–44.

[80] Hasin, "From 'Space Law' to 'Space Governance,'" 413.

[81] Nasr et al., "A Policy Framework for Sustainable and Equitable Space Resource Utilization," XXXX.

[82] Masson-Zwaan and Sundahl, "National and International Norms towards the Governance of Commercial Space Resource Activity," 402.

[83] Jakhu and Buzdugan, "Development of the Natural Resources of the Moon and Other Celestial Bodies," 228–32; Moltz, "Toward Cooperation or Conflict on the Moon?," 100; Hasin, "From 'Space Law' to 'Space Governance,'" 430; Masson-Zwaan and Sundahl, "National and International Norms towards the Governance of Commercial Space Resource Activity," 399–402.

[84] Sundahl and Masson-Zwaan, "The Lunar Legal Landscape," 54.

The trends in diplomacy under polycentrism and literature review helped to highlight the space diplomacy mechanisms outlined by Cross and Pekknanen[85] and four scenarios for space governance by Yap et al.[86] as relevant approaches for operationalization of the stated research aim and further qualitative sorting of the collected data.

The model for governing commons in space introduced by Yap et. al, in 2023 proposes four alternative scenarios. The specific scenarios include Earth-space sustainability, Open space, Earth-centric gold rush, and Space cartel.[87] Using space diplomatic mechanisms to operationalize these scenarios led to an underlining of the activities of scientific research networks. The networks aim to collaborate and reach shared goals. This tendency contrasts with governmental actors who pursue the fulfillment of national self-interests[88]. On the level of diplomatic negotiations, these dichotomies open space for operationalizing the four alternative scenarios for predicting possible negotiation outcomes.

*Earth-space Sustainability*

In the Earth-space sustainability scenario, the ideas of a space ecological epistemic community promoting environmental protection of outer space are the most relevant.[89] Its concepts shape and facilitate an information exchange within groups of diplomats and epistemic communities. Also, it frames outreach initiatives for the general public toward accomplishing a shared sustainability goal for the Earth and outer space.[90] The scenario introduces sustainability as a complex and relevant issue and the Earth as a non-isolated space object. The sustainability concept is embedded in 'diplomacy for science'[91] policies to shape the top-down research priorities and bottom-up collaboration between individual researchers and their interactions with political and diplomatic officials. Implementing diplomacy for science policies involves negotiating and reaching agreements supporting participation in multinational programs and building major research infrastructures.[92] In these activities, the business sector emphasizes the philanthropist role as it shares benefits from outer space across regions on Earth. Therefore, science communicators frame space activities as opportunities for solutions to present-day challenges such as mitigating the effects of climate change and minimizing environmental degradation. As this scenario emphasizes a fair distribution of space resources among governmental and non-governmental actors across countries at different developmental stages, it stimulates diversity and inclusion of global audiences toward space activities. Moreover, the competition among countries is less intense, so it lowers the developmental gap between developed and developing countries, improves coordination, and provides equal access to space resources. Countries with a status of 'followers'[93] have guaranteed access to the means for space activities and space resources, while achieving democracy in their political systems represents the top priority.

[85] Cross and Pekkanen, "Introduction. Space Diplomacy," 202–6.
[86] Yap et al., "Four Alternative Scenarios of Commons in Space," 397–402.
[87] Yap et al., "Four Alternative Scenarios of Commons in Space," PP
[88] Cross and Pekkanen, "Introduction. Space Diplomacy," 204.
[89] Trur, "Governance Aspects of Space Sustainability," 453; Cross, "Outer Space and the Idea of the Global Commons," 386.
[90] Cross and Pekkanen, "Introduction. Space Diplomacy," 202–3.
[91] The Royal Society, "New Frontiers in Science Diplomacy - Navigating the Changing Balance of Power," 9.

*Open Space (Space Utopia)*

In the open space model, actors operate in multiple multi-level governance settings. This model encompasses global, regional, national, and local levels. A mixture of larger and smaller business ventures replaces the exclusivity of multinational corporations and implies double regulatory requirements hindering innovation and commercialization. The model aligns with recent trends in environmental governance, which assumes decentralization and localization of agendas. Therefore, diverse space actors consider various environmental and societal factors, including location, economic development, and specific knowledge discourse. The open space governance model itself indicates fragmentation from an external perspective. Internally, the diversity of actors and their wants establishes cross-scale communication as a priority to use a market-driven rational choice for avoiding the tragedy of space commons to secure the sustainability of their space activities.[94] [95] Although the open space model provides opportunities for high competition among actors, sustainability as a prevalent scientific paradigm provides a basis for building trust among

[92] Ruffini, *Science and Diplomacy A New Dimension of International Relations*, 12.
[93] Yap et al., "Four Alternative Scenarios of Commons in Space," 401–2.
[94] Yap et al., 400–401.
[95] Patman and Davis, "Science Diplomacy in the Indo-Pacific Region," 865.

the diverse actors.[96] [97] It makes their actions confidential. Together with a need to solve global challenges being out of traditional governmental control and a prevalent narrative highlighting science's ability to rationalize and pacify egoistic self-interest in politics via providing complex information, the trust among diverse actors in the open space model allows spillover of confidentiality into non-scientific dimensions of politics to provide a foundation for building or improving diplomatic relations, and fostering cross-border policy integration and international policy collaboration.[98] [99]

*Earth-centric Gold Rush*

The Earth-centric gold rush scenario and the space cartel scenario share an assumption. In both, large multinational companies are the primary actors responsible for utilizing outer space and have strong financial, political, and legal support from the governments of space-faring countries. The difference between scenarios lies in framing associated with persuasion targeting changing and uniforming the attitudes of involved actors.[100] [101] Therefore, this diplomatic strategy has the ability to provide more profound knowledge for a more persuasive argument to change the frame of understanding of space exploration for facilitating diplomatic negotiations and reaching a consensus on a new set of global values.[102] While the space cartel scenario frames the activities of multinational private actors as the most crucial aspect to accelerate accomplishing scientific and technological progress and narrows their regulation to non-binding best practices, the Earth-centric gold rush scenario uses the mitigation of environmental degradation within national territories as shared norms, principles, history and motivation to appropriate use of the space resources. So, mitigation of transboundary environmental degradation provides new aspects for a preferred way of thinking, casual interpretation, and moral evaluation to serve as minilateral agreements, which raises the predictability and manageability of policy consequences stemming from the implementation of this scenario.[103] However, from a future perspective, minilateralism indicates a high probability of tragedy for the outer space commons.[104]

---

[96] Flink, "Taking the Pulse of Science Diplomacy and Developing Practices of Valuation," 192–94.
[97] Flink, 192–94.
[98] Gore, Nichols, and Lips, "Preparing Scientists for Science Diplomacy Requires New Science Policy Bridges," 430.
[99] Su and Mayer, "Science Diplomacy and Trust Building," 24, 26; Rungius and Flink, "Romancing Science for Global Solutions," 2,4.
[100] Provis, "Negotiation, Persuasion and Argument," 99; Morales and Berridge, "'Liking' and 'Wanting' in Eating and Food Reward."
[101] Bosse, Höpner, and Vieira, "Persuading the Giant?," 299–300; Grobe, "The Power of Words," 12–13.
[102] Provis, "Negotiation, Persuasion and Argument," 99–100.

*Space Cartel*

In the space cartel scenario, national governments in leading space faring countries delegate responsibilities for implementing outer space activities to technologically advanced private actors due to their innovative potential. Governments prioritize establishing public-private partnerships, prioritizing large private companies as the most relevant actors for outer space's use and exploration. Public-private partnerships respect the conceptualized national interest. So, their activities aim to avoid costly domestic or international conflicts and protect resources within national borders from internal and external threats. The exclusiveness of public-private partnerships opens doors for private actors who represent governmental sovereigns and provide a basis for their hegemonic aspirations.[105] *"Strategic negotiations result in tit-for-tat trade-offs among space diplomats (state or non-state) who seek to maximize their state's self-interest through bargaining with other states."* [106] Private actors' activities follow the 'first-come-first-served'[107] principle for using and occupying space resources. The growing role of private actors would lead to fragmentation of international coordination and slow down negotiations about a new comprehensive regulative instrument for global space activities. Instead, international diplomatic negotiations will aim for specific, self-coordinated, and non-binding best practices aligning with private actors' interests, which would marginalize principles for global justice and sustainable development and their enforcement.[108] Private actors lead diplomatic bargaining aiming to promote national policy objectives in international arenas.[109] The range of policy implications and benefits stemming from promoting a state's foreign policy interests may increase if governmental actors use epistemic motivation and social value justification, having reciprocal fairness at its core.[110]

[103] Dekker and Scholten, "Framing the Immigration Policy Agenda," 202–4.

[104] Yap et al., "Four Alternative Scenarios of Commons in Space," 399–400.

[105] Pasquali, "The Role of Diplomatic Culture in International Disputes," 277–78; Chattu et al., "Global Health Diplomacy, Health and Human Security," 2; Keys, "The Diplomat's Two Minds," 21; Jönsson, "Theorising Diplomacy," 24.

[106] Cross and Pekkanen, "Introduction. Space Diplomacy," 204–5.

[107] McKeown, Dempster, and Saydam, "Artemis Accords," 14.

[108] Yap et al., "Four Alternative Scenarios of Commons in Space," 397–99.

[109] Hart and Siniver, "The Meaning of Diplomacy," 13–14; Rathbun, Kertzer, and Paradis, "Homo Diplomaticus," S40–41.

## The Alternative Scenarios Applied to The Working Group Negotiations

### *Space Cartel*

In the space cartel scenario context, the delegations of Russia, Belarus, and China recognized the innovative potential and financial resources of technologically advanced actors in outer space research. Concurrently, these delegations called for substantive attention to elaborate the non-appropriation and benefit-sharing principles within the future legal regime for space resources activities. The prospective debate in the Working Group should prevent the first-come-first-served principle for using and occupying space resources and maintain the international status of space resources.[111] The delegations outlined the consequences of the limited volume and location of space resources as the limitations can stimulate economic rationality and the self-centered behavior of technologically advanced space actors.[112] Especially China set on the agenda a demand for an effective governmental supervision over space resources activities by non-governmental entities, especially by private actors. The supervision should prevent hegemonic aspirations of these non-governmental actors.[113]

These undesirable tendencies in the context of the Outer Space Treaty would slow negotiations for subsequent diplomatic talks about a new comprehensive regulatory instrument for global space activities and contribute to the undesirable fragmentation of the international community of policymakers. So, especially the Russian submission highlighted a need for maintaining the international status of space resources after their extraction and their subsequent transformation to

110 Rathbun, Kertzer, and Paradis, "Homo Diplomaticus," S40–41.
111 UNOOSA, "A/AC.105/C.2/2023/CRP.20 - Russian Federation – Input to the Working Group on Legal Aspects of Space Resource Activities," 5–6.
112 UNOOSA, 7.
113 UNOOSA, "A/AC.105/C.2/2024/CRP.5 - China – Input to the Working Group on Legal Aspects of Space Resource Activities," 3.

consider the level of economic and scientific development to ensure the fair distribution of benefits from space mining activities.[114]

Regarding the level of observer inputs to the Working Group, of the seven observer organizations, Open Lunar Foundation's (OLF) contribution to the debate exhibits the most concern about the known challenges of common pool resource issues such as scarcity, rivalry, and developing sustainable polycentric governance mechanisms.[115] Input from Space Generation Advisory Council (SGAC) differs in significant ways from OLF, where OLF is defining and categorizing types of resources and predetermining scarcity and rivalry of Lunar resource systems with a potential for private property (policy permitting),[116] the SGAC, otherwise advised that currently the "extent to which off-Earth resources can contribute to public and private space activities cannot be exactly determined due to the lack of sufficient and reliable data on their availability and the economic viability of their extraction and use."[117] In this regard, SGAC asserted that "what constitutes a highly valuable or limited resource is likely to evolve as more data becomes available, and so will the related need to regulate their access and use." Moreover, SGAC contend that "the concept of what constitutes a *lunar resource* and the legal conditions for their access and use should evolve side by side."[118]

*Open Space (Space Utopia)*

Regarding the open space model, Luxembourg, the Netherlands, Belgium, and other Western countries submitted proposals to spark the debate about multiple multi-level governance settings. The Building Blocks for the Development of an International Framework on Space Resource Activities for which the outputs of the achievements of the International Seabed Authority may serve as a good practice constituted the core of their positions.[119] Considering the interests and benefits of all countries, the countries' proposals recognized the diversity of space actors with various socio-economic and environmental needs. Therefore, the Building Blocks should be supportive for legal certainty, cross-scale communication, and mitigate the risk of international conflicts. These aspects stimulate the roles and needs of commercial actors for whom market-driven mechanisms help to avoid the tragedy of space commons.

[114] UNOOSA, "A/AC.105/C.2/2023/CRP.20 - Russian Federation – Input to the Working Group on Legal Aspects of Space Resource Activities," 7.
[115] UNOOSA, "A/AC.105/C.2/2023/CRP.25 - Open Lunar Foundation – Input to the Working Group on Legal Aspects of Space Resource Activities," 10–11.
[116] UNOOSA, 8.
[117] UNOOSA, "A/AC.105/C.2/2023/CRP.26 - Space Generation Advisory Council – Input to the Working Group on Legal Aspects of Space Resource Activities," 49.
[118] UNOOSA, 49.

Moreover, the benefit-sharing principle for utilizing space resources ensures access and exchange of information and might result in the establishment of joint ventures. The introduced benefit-sharing principle may require the establishment of an international fund for voluntary contributions, especially from the space operators responsible for conducting space resource activities. The safety zones introduced in the Artemis Accords have potential to reduce uncertainty and specify the critical support for safe and sustainable operations.[120] In sum, the submitted proposals from Western countries equally predicted an emergence of decentralized and localized agendas on space resources that may result in a form of adaptive governance.

At the observer level, Western ideas predominated (see: Appendix 1.), we evaluated that at this stage of the Working Group, all observer organizations are based in the United States and Europe. Of these, Moon Village Association (MVA), SGAC, and OLF, were most vocal about aspects of adaptive governance. MVA views the challenge for COPUOS as not only to adopt a legal model for space resources, but to develop an adaptive legal model with sustainable rules for future generations.[121] Similarly, OLF views the development of adaptive governance mechanisms as necessary to understanding how to recover, process and use space resources safely and sustainably.[122] In this regard, both MVA and OLF prioritized development enabled by adaptive governance mechanisms.

[119] UNOOSA, "A/AC.105/C.2/L.315 - Building Blocks for the Development of an International Framework on Space Resource Activities: Working Paper Submitted by Luxembourg and the Netherlands," 4; UNOOSA, "A/AC.105/C.2/L.325 - Contribution by Belgium to the General Exchange of Views on Potential Legal Models for Activities in the Exploration, Exploitation and Utilization of Space Resources: Working Paper Submitted by Belgium," 3.

[120] UNOOSA, "A/AC.105/C.2/2023/CRP.7 - Australia – Input to the Working Group on Legal Aspects of Space Resource Activities," 6.

[121] UNOOSA, "A/AC.105/C.2/2023/CRP.23 - Moon Village Association – Input to the Working Group on Legal Aspects of Space Resource Activities," 2.

Of the seven observers, concerns over benefit sharing were most pronounced in SGAC, MVA, National Space Society (NSS) and European Space Agency (ESA) inputs. The NSS and ESA emphasized the importance of clarifying benefit-sharing, especially in the legal field and what it means for incentivizing investment, the development of technology, capabilities, and education, including for developing countries.[123] MVA called for a "broad and inclusive debate on measures to mitigate future inequalities that may result from lack of sufficient consideration of access and benefit sharing issues" with the view that this theme should be "mandatory at relevant legal and operational levels," and this must be preceded by resolving access issues in "close concertation with key operators and strongly invested stakeholders".[124]The SGAC with regard to a future Lunar Governance Charter, should include an Annex protocol for internationally agreed benefit sharing and capacity building based on principles of adaptive governance.[125] Still with observer views regarding adaptive governance and benefit sharing priorities in the Working Group's deliberations, it does not guarantee any level of equal access or share of future benefits but could potentially evolve towards this scenario.

*Earth-Space Sustainability*

Although the aspects of Earth-space sustainability were marginal in the submitted proposals, some European countries and New Zealand considered space resource utilization a novel domain for space exploration with uncertain consequences for the Earth and outer space environments. They expressed their interest in establishing an epistemic community to evaluate various aspects of accomplishing shared sustainability for Earth and outer space.[126] Also, future negotiations in the Working Group should build a consensus on responsibilities for damage to the environment, including legal instruments for the prevention of adverse changes in the Earth's environment.[127]

---

122 UNOOSA, "A/AC.105/C.2/2023/CRP.25 - Open Lunar Foundation – Input to the Working Group on Legal Aspects of Space Resource Activities," 5.

123 UNOOSA, "A/AC.105/C.2/2023/CRP.24 - National Space Society – Input to the Working Group on Legal Aspects of Space Resource Activities," 3; UNOOSA, "A/AC.105/C.2/2023/CRP.22 - European Space Agency – Input to the Working Group on Legal Aspects of Space Resource Activities," 3.

124 UNOOSA, "A/AC.105/C.2/2023/CRP.23 - Moon Village Association – Input to the Working Group on Legal Aspects of Space Resource Activities," 3–4.

125 UNOOSA, "A/AC.105/C.2/2023/CRP.26 - Space Generation Advisory Council – Input to the Working Group on Legal Aspects of Space Resource Activities," 7.

Notably, the UK delegation highlighted taking into account COSPAR's recommendations for planetary protection policies.[128] The expert panel elaborated five categories to operationalize potential hazards from space exploration for the Earth's environment.[129] The categories range from cases requiring no protection up to missions having a status needing continuous control throughout the return to Earth phase.[130] Following the expected risks, COSPAR's panel called for reporting information on space missions that considered the concept of planetary protection. The content of these reports should provide detailed data on biological burdens.

Observer inputs to the working group have used the term "sustainability" with regard to space resource activities multiple ways referring to a variety of issues whether this involves the sustainability of a legal framework, sustainability of projects and human missions, or building and sustaining economic aspects in space and on Earth. Yet few, if at all, involve sustainability in an environmental sense. The National Space Society prioritized protecting heritage sites in their statement but did not separate this from the distinct complexities of interference aspects.[131] For those fitting aspects of an Earth-Space sustainability scenario, the NGO For All Moonkind viewed Space resource utilization as an absolute necessity and advocates for enabling resource use while dealing with protecting "cultural heritage" sites. Moreover, they view the resources of space as vast and have the potential to better the life of our human species both on Earth and in the Universe.[132] Open Lunar Foundation view "the Moon as part of a developing Earth–Moon system"[133] and the Space Generation Advisory Council asserted that governmental funding should be complemented by dedicated measures to promote deeper integration with Earth's economy and industries. Therefore, funding a private lunar economy would democratize access to the Moon and reduce pressure upon States.[134] Of all the observer inputs, the European Space Agency space resources strategy aligns most with this scenario declaring that the challenge of space resources will also drive innovation on Earth.[135] According to ESA, learning to use the resources of space creates opportunities for Earth through technology innovation and new approaches to managing the resources of Earth. In this regard, ESA views the challenges and environments of space as "driving more optimized technologies and systems with implications for sustainability and the Sustainable Development Goals and by addressing the increasing challenge of resource scarcity where the resources of space may even be used on Earth as the cost of access falls and some resources on Earth become increasingly scarce."[136]

---

[126] UNOOSA, "A/AC.105/C.2/2023/CRP.18 - New Zealand – Input to the Working Group on Legal Aspects of Space Resource Activities," 3; UNOOSA, "A/AC.105/C.2/2022/CRP.21* - Working Paper on the Endorsement of the Work Plan for the Working Group Established under Agenda Entitled 'General Exchange of Views on Potential Legal Models for Activities in the Exploration, Exploitation, and Utilization of Space Resources' and Proposals for a Dedicated International Conference: Working Paper Submitted by Austria, Belgium, Czech Republic, Finland, Germany, Luxembourg, Norway, Portugal and Romania," 3.

[127] UNOOSA, "A/AC.105/C.2/L.328 - Updated Summary by the Chair and Vice-Chair of Views and Contributions Received on the Mandate and Purpose of the Working Group on Legal Aspects of Space Resource Activities," 2; UNOOSA, "A/AC.105/C.2/L.315 - Building Blocks for the Development of an International Framework on Space Resource Activities: Working Paper Submitted by Luxembourg and the Netherlands," 4; UNOOSA, "A/AC.105/C.2/2023/CRP.13 - Germany – Input to the Working Group on Legal Aspects of Space Resource Activities," 3; UNOOSA, "A/AC.105/C.2/2023/CRP.14 - Greece – Input to the Working Group on Legal Aspects of Space Resource Activities," 5.

[128] UNOOSA, "A/AC.105/C.2/2023/CRP.21 - United Kingdom – Input to the Working Group on Legal Aspects of Space Resource Activities," 3.

[129] COSPAR, "Committee on Space Research (COSPAR) » Panel on Planetary Protection (PPP)."

[130] COSPAR, "COSPAR Policy on Planetary Protection," 1–3.

[131] UNOOSA, "A/AC.105/C.2/2023/CRP.24 - National Space Society – Input to the Working Group on Legal Aspects of Space Resource Activities," 3,5.

*Earth-Centric Gold Rush*

In this scenario, countries' proposals indicated the lack of data needed to evaluate future business behavioral practices and business actors' internal motivations. At the declaratory level, the proposals underlined existing relevant international legal principles that should shape the subsequent utilization of outer space, such as benefits-sharing, considering outer space as the province of all humanity, and minimizing negative externalities for Earth's environment.[137] Moreover, the working group should elaborate on the responsibilities of strengthening conflict resolution mechanisms among actors participating in the extraction and utilization of space resources.[138]

---

[132] UNOOSA, "A/AC.105/C.2/2023/CRP.35 - For All Moonkind – Input to the Working Group on Legal Aspects of Space Resource Activities," 5.

[133] UNOOSA, "A/AC.105/C.2/2023/CRP.25 - Open Lunar Foundation – Input to the Working Group on Legal Aspects of Space Resource Activities," 4.

[134] UNOOSA, "A/AC.105/C.2/2023/CRP.26 - Space Generation Advisory Council – Input to the Working Group on Legal Aspects of Space Resource Activities," 37.

[135] ESA, "ESA Space Resources Strategy," 6.

[136] ESA, 8, 11.

A Belgian contribution to the debate in the Working Group conceptualized and framed risks that could lead to the tragedy of outer space commons.[139] Although unilateral and club-based policy and legal instruments can potentially increase the efficiency of project-driven tasks, the approach is not in line with existing Space Treaties and may deepen economic inequalities worldwide. The Belgian delegation welcomed future development of forecast models for space resource utilization. It also emphasized responsibilities for private actors involved in space mining activities to avoid resource exhaustion. The multilateral global dialogue could facilitate information exchange, create conditions for transparency, and build confidence to reduce the risk of conflicts.[140] Also, future negotiations should discuss establishing an international fund for efficient benefit sharing.[141]

All observer inputs to the Working Group in subtle ways portend the possibility of aspects of both the Earth centric gold rush and space cartel scenarios in that private actors are critical to scientific and technological progress and require legal certainty. Input from the European Space Agency prioritized public private partnerships[142] while the Hague Institute for Global Justice asserted that the private sector needs a "global framework that is transparent, uniform, simple, and predictable" and "achieved through a bottom-up approach that relies on the advice and experience of civil society."[143] Additionally, the SGAC highlighted the development of a lunar economy as an essential component, finding that lunar companies would benefit from support of the public sector where governments and space agencies should provide assistance for non-governmental lunar activities.[144] The NGO For all Moonkind asserted that to avoid excessively high costs of space exploration both States and private companies "must be able to extract, use and create commerce around space resources" in their view it is expected and must be "permitted that space resources will be able to be used to supplement dwindling natural resources on Earth."[145]

---

[137] UNOOSA, "A/AC.105/C.2/2023/CRP.18 - New Zealand – Input to the Working Group on Legal Aspects of Space Resource Activities," 1–3.
[138] UNOOSA, "A/AC.105/C.2/2023/CRP.20 - Russian Federation – Input to the Working Group on Legal Aspects of Space Resource Activities," 2–4.
[139] UNOOSA, "A/AC.105/C.2/L.325 - Contribution by Belgium to the General Exchange of Views on Potential Legal Models for Activities in the Exploration, Exploitation and Utilization of Space Resources: Working Paper Submitted by Belgium," 2–3.
[140] UNOOSA, "A/AC.105/C.2/2023/CRP.13 - Germany – Input to the Working Group on Legal Aspects of Space Resource Activities," 3; UNOOSA, "A/AC.105/C.2/2023/CRP.37 - United States – Input to the Working Group on Legal Aspects of Space Resource Activities," 4–5.
[141] UNOOSA, "A/AC.105/C.2/L.315 - Building Blocks for the Development of an International Framework on Space Resource Activities: Working Paper Submitted by Luxembourg and the Netherlands," 5; UNOOSA, "A/AC.105/C.2/2023/CRP.12 - France – Input to the Working Group on Legal Aspects of Space Resource Activities," 5–6.
[142] ESA, "ESA Space Resources Strategy," 11.

## Discussion

The results of the presented qualitative analysis indicate a diversity of views about policy and legal aspects of space resource utilization. They highlight aspects of the discussions about future governance and legal frameworks for using resources and exploring near-Earth outer space, emphasizing the lunar environment. In sum, general agreement about the need for a comprehensive clarification of hard and soft legal principles governing recent uses of outer space was visible across the submitted proposals by countries and observer organizations. In general, setting this topic on the legal subcommittee of COPUOS renewed scholar discussion about how to continue the subsequent developments in regulatory aspects of lunar activities.

Our analysis captured a variety of demands on the country level indicating that negotiations at the Working Group are still in the introductory phase. Developments at the Working Group should clarify whether and what historical international governance models, namely the Antarctic governance regime and the high seas and seabed international governance regimes, and what their particular elements are eligible to be included in desired lunar governance principles.[146] Although the international situation of the 1950s allowed the establishment of the Antarctic governance regime as an example of a temporary conflict prevention situation, in 2009 Moltz had noticed that freezing territorial claims in outer space has minority support among policy and legal scholars.[147] Our results also confirmed these tendencies at the level of international political negotiations. In the space cartel scenario, where fears about resource exhaustion were the most noticed, neither countries nor observers demanded freezing space mining activities. Instead, countries and observers synchronously called for fair sharing of benefits and maintaining their international status of the resources after extraction from outer space objects if necessary. At the level of observers, the NGO For all Moonkind outlined a hierarchy to support efficient space resource management.

[143] UNOOSA, "A/AC.105/C.2/2023/CRP.32 - The Hague Institute for Global Justice: The Washington Compact on Norms of Behavior for Commerical Space Operations," 4.
[144] UNOOSA, "A/AC.105/C.2/2023/CRP.26 - Space Generation Advisory Council – Input to the Working Group on Legal Aspects of Space Resource Activities," 37–39.
[145] UNOOSA, "A/AC.105/C.2/2023/CRP.35 - For All Moonkind – Input to the Working Group on Legal Aspects of Space Resource Activities," 5.
[146] Mouat et al., "(Un)Earthly Governance," 133.

Therefore, the debate on equitable uses of lunar resources resulting in setting an environment without significant political and economic conflicts between spacefaring and non-space faring countries became more profound instead. On the country level, actors' perspectives converge into two scenarios mainly. Most non-Western countries called for preventative measures against applicability of the first-come-first-served principle for the cases of space resource utilization. In their proposals, these countries warned against growing global inequalities and the chain of events leading toward a tragedy of the commons. Conversely, Western countries clustered their visions of lunar governance around the space utopia scenario in which emphasis on liberal values was visible. In a nutshell, they prioritized power dispersion among actors with various socio-economic and environmental needs for whom principles of cross-scale communication and market-driven mechanisms became crucial principles.

In line with Western countries' demands, the prevalent scholar discourse highlighted the inefficiency of the Outer Space Treaty for developing commercial aspects of space exploration, mainly in connection with lunar resources.[148] As space lunar resource governance remains undeveloped, the UN system should support academic and political events dealing with various aspects of excludability and rivalry.[149] These aspects should include development of advanced mechanisms for environmental impact assessments.[150] The scholarly demands were in line with the obtained results in our analysis in which countries and observers showed respect for the existing provisions of the Outer Space Treaty and further called their next elaboration to enhance legal and technological certainty.

[147] Moltz, "Toward Cooperation or Conflict on the Moon?," 87.
[148] Moltz, 94.
[149] Krolikowski and Elvis, "Space Resources and Prospects for Contested Governance," 671.
[150] Martin and Freeland, "Back to the Moon and Beyond," 428.

## Conclusion

This analysis mapped the diversity of demands of countries and observers at the Working Group to provide preliminary results about international negotiations on future governance models for space resource utilization. The obtained qualitative data indicated the progress of the ongoing expert and diplomatic talks, whose complexity is expected to increase. Their future outcomes have the potential to provide an innovative variety of governance tools for exploiting lunar resources which the existing governance regimes for global climate and oceans may inspire.[151] Also, some complex proposals for regulating radio frequencies may contribute to increasing the efficiency of the governance of limited lunar resources.[152]

The presented results showed a need for more structure in political debates in COPUOS. Future negotiations in the Working Group should cover at least two essential aspects - feasibility and effectiveness.[153]In terms of feasibility, previous research has noted the rapidly changing global environment and the differing positions of Western democratic and non-Western countries. The differences in political regimes in these two blocks portend the mutually opposite values at the UN level.[154] Whereas space activities have been the domain of governments up until recent times, current policy and legal environments in Western democratic countries open doors for private actors' involvement. These developments raise questions about the matrix of governmental, hybrid, and private responsibilities and their consequences in competing environments between Western and non-Western countries. These differences in values would be visible in framing private actors' activities in the lunar environment, especially in the structure of public-private partnerships enabling governmental incentives for these new space players.[155] Regarding the aspect of effectiveness, Neto highlighted managing space diplomacy negotiations toward widely respected solutions.[156] The ability to persuade reluctant actors to remain involved in negotiations increases relevance, especially in enabling a sustainable environment for space activities and securing the natural environment on Earth. In the context of the recognized diversity of actors existing in the space domain, building conditions toward adaptive space governance could contribute to finding a policy consensus in multilateral space governance fora.[157]

[151] Hasin, "From 'Space Law' to 'Space Governance,'" 398–99.
[152] Masson-Zwaan and Sundahl, "National and International Norms towards the Governance of Commercial Space Resource Activity," 400–401; Roberts and Linares, "A Method for Assessing Satellite Operators' Compliance with Geosynchronous Orbital Assignments," 226–28.
[153] Hasin, "From 'Space Law' to 'Space Governance,'" 409.
[154] Hasin, 408.
[155] Jakhu and Buzdugan, "Development of the Natural Resources of the Moon and Other Celestial Bodies," 206.

Future research could consider overcoming limitations that emerged during our research process. Dealing with a newly established working group at the UN and a rapidly evolving topic with limited data input thus far allowed us to provide preliminary outputs only. As the process continues and negotiations evolve, future research should consider conducting interviews with distinguished individuals and stakeholders. Future studies should also consider doing a cost-benefit analysis for particular governance models expected in the Working Group in the future.

[156] Neto, "Outer Space as a Global Commons and the Role of Space Law," 13.
[157] Oltrogge and Christensen, "Space Governance in the New Space Era," 434.

## Appendix 1: Submissions from state members and permanent observer organizations to the Working Group as of 30 August 2024

| Actor type | Name | Submission year |
|---|---|---|
| **State members** | Australia | 2022 |
| | Azerbaijan | 2022 |
| | Bahrain | 2022 |
| | Belarus | 2022 |
| | Belgium | 2023 |
| | Canada | 2023 |
| | China | 2024 |
| | France | 2022 |
| | Germany | 2022 |
| | Greece | 2022 |
| | Japan | 2023 |
| | Jordan | 2022 |
| | Luxembourg | 2022 |
| | Morocco | 2022 |
| | New Zealand | 2023 |
| | Norway | 2022 |
| | Russia | 2022 |
| | United Kingdom | 2022 |
| | United States | 2022 |
| **Permanent observer organizations** | European Space Agency[158] | 2022 |
| | For All Moonkind[159] | 2022 |
| | Hague Institute for Global Justice[160] | 2023 |
| | Moon Village Association[161] | 2022 |
| | National Space Society[162] | 2022 |
| | Open Lunar Foundation[163] | 2022 |

[158] The European Space Agency is a governmental organization based in Paris, France.
[159] The For All Moonkind is a non-governmental organization based in New York, the U.S.
[160] The Hague Institute for Global Justice is a non-governmental organization based in the Hague, Netherlands.
[161] The Moon Village Association is a non-governmental organization based in Vienna, Austria.
[162] The National Space Society is a non-governmental organization based in Washington, D. C., the U.S.
[163] The Open Lunar Foundation is a non-governmental organization based in San Francisco, the U.S.

| | | |
|---|---|---|
| | Space Generation Advisory Council[164] | 2022 |

Source: UNOOSA. Working Group on Legal Aspects of Space Resource Activities. available at: https://www.unoosa.org/oosa/en/ourwork/copuos/lsc/space-resources/index.html, last visited: 30 August 2024.

164 Space Generation Advisory Council is a non-governmental organization based in Washington, D.C., the U.S. and Vienna, Austria.